# Towards Comparing Programming Paradigms

Igor Ivkic, Markus G. Tauber
University of Applied Sciences Burgenland
Eisenstadt, Austria
e-mail: {igor.ivkic,markus.tauber}@fh-burgenland.at

Alexander Wöhrer
University of Vienna
Vienna, Austria
e-mail: alexander.woehrer@univie.ac.at

*Abstract*—Rapid technological progress in computer sciences finds solutions and at the same time creates ever more complex requirements. Due to an evolving complexity today's programming languages provide powerful frameworks which offer standard solutions for recurring tasks to assist the programmer and to avoid the re-invention of the wheel with so-called "out-of-the-box-features". In this paper, we propose a way of comparing different programming paradigms on a theoretical, technical and practical level. Furthermore, the paper presents the results of an initial comparison of two representative programming approaches, both in the closed SAP environment.

## I. Introduction

Programming may be considered by some as an art form and/or by others as a craftsmanship, but it leaves little room for discussion about its incredibly fast development. Whereas in the past developers implemented classic desktop programs, today's applications require world-wide connectivity, web presence and mobile assistance. In many cases a technological leap is followed by a change in the method (or the way the technology is used). While in the past most problems were solved with a structured programming paradigm and a data-driven approach, the new frameworks require an object oriented (OO), generic and model-driven approach [1].

Although frameworks promise to standardize program code, save development time and costs they are often caught in the crossfire of criticism due to the obscure relationship between their pains and gains. The aim of this short paper is to present an approach of how to create a full-comparison of two programming paradigms, namely ABAP - (Advanced Business Application Programming) a 4th Generation Programming Language and BOPF – (Business Object Processing Framework) a modular framework which provides custom services and components, both in the closed SAP environment. In this regard, a comparison is defined as full, when theoretical, technical and practical differences of two programming approaches were identified. So, the idea behind splitting up the approach in these three areas was to look at two software development paradigms from many different angles to create a full-comparison and to, ultimately, paint the big picture.

There are three reasons for choosing two SAP approaches for an initial demonstration of the presented approach. First, SAP is a closed system which makes a technical and practical comparison easier. Second, in 2012 SAP released the Business Object Processing Framework (BOPF) which is an extension of the existing Advance Business Application Programming (ABAP) language. This means, that SAP built and released a new model-driven framework with powerful out-of-the-box features - programmed with ABAP. Third, due to a lack of documentation and nonexistent literature there were many questions regarding the differences of these two programming approaches within the SAP programming community.

In this short paper, we present an approach for a full-comparison of two programming approaches (ABAP/BOPF) in a closed system (SAP) to identify theoretical, technical and practical differences. First, a literature research provides process- and program-theoretical differences of ABAP/BOPF. Next, for the technical comparison, a performance analysis measures the execution time of CREAT, READ, UPDATE, DELETE (CRUD) operations in an ABAP report compared to a BOPF report. Finally, a use case driven experimental study including a post-experiment User Experience Questionnaire (UEQ) [2] provides practical differences of ABAP and BOPF.

The rest of the paper is organized as follows: Section II summarizes related work in the field, followed by a technical performance analysis in Section III. Section IV deals with an empirical user experience study and a Post-Experiment Survey.

## II. Related Work

The comparison of products, services, programming languages, etc. was the subject of many papers. Most of these papers i.e.: compare a specific aspect of a product and present which product was better, but none compares them on a theoretical, technical and practical level. In example, in [3] the BOPF was firstly introduced to the public and compared to ABAP just on a technical level in a "How-To"-manner. Another comparison in [4] used the UEQ to measure UX in interactive products. Related work in [5] comes the closest, but the research field was restricted to mobile development.

## III. Performance Analysis Between Approaches

Even though BOPF was programmed with ABAP the approaches could not be more different. While developing with ABAP goes hand in hand with classic models (i.e.: waterfall model, V-model) the BOPF welcomes agile development. Following the agile principles, the BOPF provides out-of-the-box tools for modelling, developing, testing and finalizing the logic of an application while involving the customer in the development process from the outset. Technically speaking, using ABAP for programming means starting from the greenfield and building an application from scratch. This greenfield-approach is a gift and a curse at the same time. For this reason, the BOPF requires first the creation of a model of the Business Object (BO) and then uses an OO Application Programming Interface (API) to control it programmatically. To be able to understand how the API works it is necessary to compare the CRUD operations.

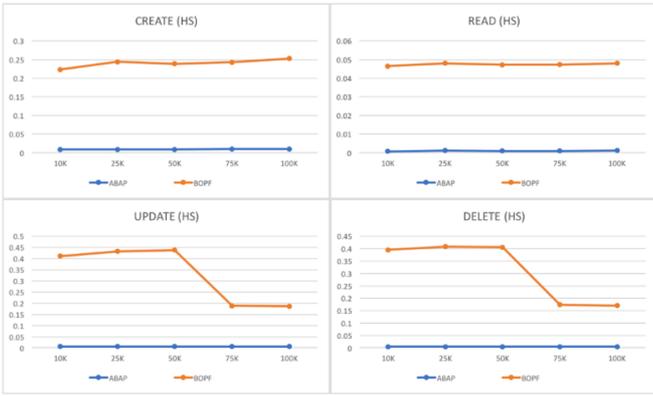

Figure 1.  Performance (execution time in ms) comparrison of CRUD operations (ABAP vs. BOPF) with increasing data block sizes

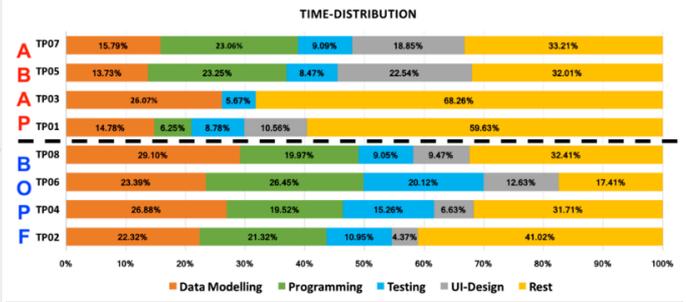

Figure 2.  Time-Distribution of the Development Process

The main difference is that each CRUD operation has its own command in ABAP while the BOPF uses two different methods for the same purpose. The MODIFY method is used to CREATE, UPDATE and DELETE records, while the QUERY method reads the records from a database. In conjunction with applications where performance plays a big role the question arises if building it with the new generic framework is the right decision. To make a direct comparison of CRUD operations and their performance on a HANA system (HS) a performance analysis was conducted. The idea was to implement eight applications (4x ABAP, 4x BOPF) where each executed one of the CRUD operations. These applications were executed one after another with five different data amounts (n = 10K, 25K, 50K, 75K, 100K). So, by choosing this approach it was guaranteed that neither the two approaches (ABAP/BOPF) nor their CRUD operations were mixed during a performance analytical run. Fig. 1 compares the results of the CRUD operations and shows the performance advantages of ABAP vs. BOPF on the HS.

IV.  DEVELOPER EXPERIENCE EXPERIMENT AND SURVEY

An experimental study was designed in two different phases which were decomposed into individual results. In the first phase eight test persons (TP) were randomly divided into two groups and confronted with a simple clear-cut programming task. The sole limitation in this scenario was that one group uses ABAP while the other one must only use the BOPF for the implementation. Building on that a survey was conducted to evaluate the user experience (UX) during the experiment. The TP answered a three-parted UEQ questioning the UX in general and how well the development environment of ABAP/BOPF supported them in solving their task. Fig. 2 shows how much time it took the TP for data modelling, programming, testing, and UI-design.

The experiment revealed that it took the BOPF group 40 minutes on average to complete 100% of the required tasks, while the ABAP group failed to implement all requirements (the TOP 2 TP finished approx. 50% of requirements after one hour; the experiment was terminated after one hour due to the superior results of the BOPF group). Finally, the post-experiment survey showed that the BOPF was rated better than ABAP by the TP. They criticized among others that programming with ABAP was inefficient and complicated in comparison to the BOPF.

V.  CONCLUSION AND FUTURE WORK

The main goal of this short paper was to present an approach which enables a systematic comparison of two software development approaches covering two aspects. In the first aspect, the comparison is done on a technical level where the performance of the program code of the two approaches was measured. The second aspect dealt with analyzing the differences of two approaches by applying them in a close to reality, use case driven experimental study.

For future work, we will develop a single metric to reflect and combine the results of all three comparisons. We will further investigate whether the results of the performance analysis and the experimental study would change when larger data amounts (n > 100K) and more than eight TP are used. Another idea for future work would be to execute the performance analysis using applications with more complex program logic (more than CRUD operations). As for the experimental study, it would also be great to find out, if the outcome of the experiment would be the same, when a more complex application is given for the implementation. Finally, it would be great to apply this approach on many different programming languages and frameworks to point out significant differences in the results.